\documentclass[runningheads]{llncs}
\usepackage{graphicx}
\usepackage{subfig}
\usepackage{cite}
\usepackage[labelfont=bf]{caption}

\usepackage[colorlinks=true, allcolors=blue]{hyperref}

\usepackage{xcolor}
\usepackage{booktabs}
\usepackage{multirow}
\usepackage{array}
\usepackage{threeparttable}

\newcolumntype{x}[1]{>{\centering\arraybackslash\hspace{0pt}}m{#1}}
\newcolumntype{z}[1]{>{\centering\arraybackslash\vspace{0.3pt}}m{#1}}
\newcolumntype{y}[1]{>{\centering\arraybackslash}m{#1}}
\newcolumntype{q}[1]{>{\arraybackslash\vspace{0pt}}p{#1}}

\begin{document}
\title{Combining Automatic Coding and Instructor Input to Generate ENA Visualizations for Asynchronous Online Discussion}
%
%
\author{Marcia Moraes\thanks{These authors contributed equally to this work.}\inst{1}\orcidID{0000-0002-9652-3011} \and
Sadaf Ghaffari$^{\ast}$\inst{1}\orcidID{0000-0002-2413-5829} \and
Yanye Luther\inst{1}\orcidID{0000-0002-0995-3356} \and
James Folkestad\inst{2}\orcidID{0000-0003-0301-8364}
\email{\{marcia.moraes, sadaf.ghaffari, yanye.luther, james.folkestad \}@colostate.edu}
}
\authorrunning{F. Author et al.}
%
\institute{Department of Computer Science\textsuperscript{1}, School of Education\textsuperscript{2}, Colorado State University, Fort Collins CO 80523, USA}
\maketitle              
\begin{abstract}
Asynchronous online discussions are a common fundamental tools to facilitate social interaction in hybrid and online courses. However, instructors lack the tools to accomplish the overwhelming task of evaluating asynchronous online discussion activities. In this paper we present an approach that uses Latent Dirichlet Analysis (LDA) and the instructor's keywords to automatically extract codes from a relatively small dataset. We use the generated codes to build an Epistemic Network Analysis (ENA) model and compare this model with a previous ENA model built by human coders. The results show that there is no statistical difference between the two models. We present an analysis of these models and discuss the potential use of ENA as a visualization to help instructors evaluating asynchronous online discussions.


\end{abstract}
\section{Introduction}
Asynchronous online discussions are a common fundamental tool to facilitate
social interaction in hybrid and online courses. They have been shown to improve students’ critical thinking~\cite{garrison2001critical}, knowledge construction~\cite{koh2010project}, writing skills~\cite{aloni2018research}, and learning outcomes~\cite{thomas2002learning}. However, instructors lack the tools to accomplish the overwhelming task of evaluating asynchronous online discussion activities~\cite{de2019expect}.
According to de Lima et al.~\cite{de2019expect}, instructors reported struggling to assess students’ contributions
in forum activities due to difficulties in following the discussions, the lack of specific reports related to the subjects discussed, the students’ contributions to those subjects, and the lack of visualizations to convey messages in a graphical format.

In order to address some of those difficulties, Epistemic Network Analysis (ENA) has been presented as learning analytics visualization tool to show the relationships among the different concepts students discuss in an asynchronous online discussion~\cite{moraes2021using}. In this particular work~\cite{moraes2021using}, two human annotators manually coded the text data. In a more recent work\cite{saravani2023automated}, three different text mining approaches, namely Latent Semantic Analysis (LSA), Latent Dirichlet Analysis (LDA) and Clustering Word Vectors, were applied to automate code extraction from a relatively small discussion board dataset obtained from \cite{moraes2021using}. Based on the study presented in \cite{moraes2021using}, we submitted a project proposal to our university’s teaching innovation grant to investigate the application of ENA as a visualization tool to support instructors. Reviewers mentioned that instructors would like to have the visualizations but would not have time to be involved in the coding process, even if that process used nCoder\cite{ncoder}. However, the instructors would be willing to provide keywords that should be present in the codes. Based on the provided feedback, and unlike previous works, we not only looked into extracting codes from relatively small data using LDA but also illustrated how this can lead to generating ENA visualization without using nCoder\cite{ncoder} throughout the process. 

This paper builds on those previous works by:
\begin{enumerate}
  \item Using the automatically generated codes provided in \cite{saravani2023automated} to code the same dataset used in \cite{moraes2021using}.
  \item Evaluating the quality of the automated coding with the human coding presented on \cite{moraes2021using} by interrater reliability.
  \item Improving the automated code generation in order to reach a satisfactory interrater reliability between algorithm and human if necessary.
  \item Applying the codes that were automatically generated  in the creation of ENA visualizations.
  \item Analyzing and evaluating those ENA visualizations with human coded ENA visualizations.
  \item Validating both ENA visualizations with instructors.
\end{enumerate}

In this paper we aim to answer the following research questions:

\begin{itemize}\item {\textbf{RQ1.} \it {What is/are the difference(s) between the ENA visualizations generated using an automated coding process and a human coding process for the data presented in}\cite{moraes2021using}?
\item \it \textbf{RQ2.} How does an instructor evaluate the ENA visualizations generated?}
\end{itemize}

The remainder of this paper is structured as follows. Section~\ref{relatedworks} summarizes prior research efforts on using ENA as a visualizations tool and on automated coding processes. Sections~\ref{methods} and~\ref{experiments} describe the approach used in the study. Section~\ref{results} is dedicated to presenting the results obtained from our experiments. We discuss the results and outline directions for further work in Section~\ref{discussion}.

\section{Related Works} \label{relatedworks}
ENA has been used to support many facets of education and learning in several areas~\cite{10.1007/978-3-030-93859-8_12, 10.1007/978-3-030-93859-8_25, 10.1007/978-3-030-93859-8_16}. One use of ENA that is gaining more attention is its use as a visualization tool~\cite{QEvisualization} to help instructors evaluate clinical team simulations\cite{10.1145/3448139.3448176}, support teachers' interventions in students' virtual collaboration~\cite{10.1145/3170358.3170394}, evaluate teamwork~\cite{swiecki2019modeling}, include participants in co-construction and co-interpretation of ENA representations\cite{vega2022}, and unveil the conceptual connections that students are making in asynchronous online discussions\cite{moraes2021using}. All of these studies utilized different coding strategies to code the data used in the ENA visualizations.

A recent work by Cai et al.~\cite{cai2021using} centered around nCoder where they investigated how close human created code words were to codes identified by topic models in two large datsets. Another work compared the performance of neural networks in a supervised learning manner with nCoder to assess which approach required the least human coding effort while achieving a sufficient and accurate classification~\cite{osti_10354410}. In their comparison, they indicated nCoder had a higher accuracy. Although nCoder is a popular learning analytics platform used to develop coding schemes, it is not fully automated. In other words, it requires active human efforts to read through every item in the data and validate if the choice of coding is conceptually valid. It also suffers the low recall problem. Previous literature studies have identified ways to improve the problem of low recall or high false negative rate in nCoder. nCoder+~\cite{cai2019ncoder+} aims to improve low recall issue in nCoder through semantic component addition. 
However, as mentioned by the authors of nCoder+, the idea is still a prototype and is not a public tool yet. In another research effort, the use of Negative Reversion Set (NRS) sampling has been shown to improve the low recall for Regular Expressions based classifiers such as nCoder~\cite{osti_10354430}. 

Our work has several distinctive features compared to prior works. First, we only take advantage of a Natural Language Processing (NLP) unsupervised learning technique to automate the code extraction, despite having a relatively small dataset without the use of nCoder. Second, we utilize coherence analysis~\cite{newman2010automatic} to identify the optimal number of topics in the discussion data, thus avoiding arbitrary selection of the number. Third, we use instructor keywords in addition to the LDA extracted keywords to generate the visualization.

\section{Method} \label{methods}
The main goal of our approach was to extract topic keywords from a relatively small online discussion dataset using Latent Dirichlet Allocation (LDA)\cite{blei2003latent}, use those keywords to automatically code asynchronous online discussion data, and generate ENA visualizations based on that data. In this section, we describe how we automated this process. 


\subsection{Dataset Preprocessing} \label{data}

The data utilized to investigate the research questions comprised of online discussion posts from seven semesters: Fall 2017, Fall 2018, Fall 2019, Spring 2020, Fall 2020, Spring 2021, and Fall 2021. The data consisted of 2,648 postings collected from an online class for organizational leaders as part of a Masters of Education program at a Research 1 land-grant university. Table~\ref{dataset} represents prior codes in our dataset. 

The problem of interest was based on code retrieval. This highlighted the importance of the preprocessing step in our setup. The preprocessing steps in our automatic extraction task consisted of tokenization, lowercasing, named-entity removal, stop words removal, in-document frequency filtering, and generating bigrams and trigrams since our interest was to retrieve the code containing two or three words.

\begin{table}[!tbh]
    
    \centering
    \setlength{\tabcolsep}{7pt}
    \resizebox{0.91\textwidth}{!}{\begin{minipage}{\textwidth}
    \caption{Priori codes}
    \begin{tabular}{m{0.2\linewidth}m{0.55\linewidth}x{0.1\linewidth}} 
    \toprule
        \textbf{Code Name} & \textbf{Definition} & \textbf{Kappa ($\kappa$)}\\
        \cmidrule{1-3}
        Retrieval practice, Spacing out practice, Interleaving
        &
        Retrieval practice is the act of recalling facts or concepts or events from memory and are also known as testing effect or retrieval-practice effect. Spacing out practice allows people to a little forgetting that helps their process of consolidation. Interleaving the practice of two or more concepts or skills help develop the ability to discriminate later between different kinds of problems and select the better solution. 
        &
        0.85\\
        \cmidrule{1-3}
        Illusion of mastery
        &
        Researches have pointed out that students usually have a misunderstanding about how learning occurs and engage with learning strategies that are not beneficial for their long-term retention, such as rereading the material several times and cramming before exams. 
        &
        0.89\\
        \cmidrule{1-3}
        Effortful learning
        &
        Learning is deeper and more durable when it is effortful, meaning that efforts, short-terms impediments (desirable difficulties), learning from mistakes, and trying to solve some problem before knowing the correct answer makes for stronger learning.
        &
        0.85\\
        \cmidrule{1-3}
        Get beyond learning styles
        &
        Researchers found that when instructional style matches the nature of the content, all learners learn better, regardless of their learning styles.
        &
        0.86\\
    \bottomrule
    \end{tabular}
    \label{dataset}
\end{minipage}}   
\end{table}

\subsection{Latent Dirichlet Analysis} \label{sub_lda}

We aimed to determine which codes are associated with each discussion post, i.e. in each document, and extract them. To accomplish this we used LDA~\cite{blei2003latent}, a generative probabilistic model, to extract the codes from the online discussion data to help understand what topics were discussed in the course. In order to find high probability words within each topic, the number of topics was set to 5 to get the high topic coherence score\cite{saravani2023automated, newman2010automatic}. Table~\ref{tab:results-lda} shows the extracted words for each topic. In Table~\ref{tab:results-lda}, \textbf{Topic 1} code words are associated with \emph{Effortful learning} code, \textbf{Topic 2} code words are associated with \emph{Get beyond learning styles}, \textbf{Topic 3} code words are associated with \emph{Illusion of mastery}, and \textbf{Topic 4} code words are associated with \emph{Retrieval practice}, \emph{Spacing out practice}, and \emph{Interleaving}. Only \textbf{Topic 0} did not represent any codes.


\begin{table} [hbt!]
\centering
\caption{Five topics extracted by Latent Dirichlet Allocation}\label{tab:results-lda}
\setlength{\tabcolsep}{3pt}
\small
\begin{tabular}{ccccc} 
 \toprule
 \textbf{Topic} 0 & \textbf{Topic 1} & \textbf{Topic 2} & \textbf{Topic 3} & \textbf{Topic 4} \\ [0.8ex] 
 \cmidrule(lr){1-5}
 lecture & desire & dylexia & confidence & mass \\ 
 solution & desire\_difficulty & learn\_style & feedback & mass\_practice \\
 classroom & plf\textsuperscript{*} & individual & calibration & interleaving\_practice \\
 surgeon & resonate & learn\_differ & confidence\_memory & space\_retrieval \\
 acquire & parachute & disable & accuracy & tend \\
 instruct & fall & intelligent & peer & day \\
 learn\_learn & land & prefer & answer & long\_term \\
 impact & jump & support & event & week \\
 demand & parachute\_land & dyslex & state & myth \\
 lecture\_classroom & land\_fall & focus & calibration\_learn & practice\_space \\ [1ex]
 \bottomrule
\end{tabular}
\begin{tablenotes}
      \footnotesize
      \item $^*$Stands for Parachute Landing Fall.
    \end{tablenotes}
\end{table}
\section{Experiments} \label{experiments}

With the LDA extracted keywords for 4 topics, we conducted experiments with those keywords alone as described in~\ref{subsec1}, and along with the keywords identified by the instructor as described in~\ref{subsec2} and~\ref{subsec3}. 

\subsection{Experiment 1}\label{subsec1}

Following the results obtained from~\ref{sub_lda}, we automatically generated a well-formatted table \cite{shaffer2016tutorial} in which each row consisted of: \emph{post entry number}, \emph{user id}, \emph{date and time for that post entry}, \emph{actual discussion post data}, and \emph{list of codes with 1's or 0's corresponding to the existence or no existence of the specific code in each post}. The table was entered into the ENA webtool in an Excel format \cite{ref_url1}. We then ran interrater reliability between the automated coding and the human coding provided by \cite{moraes2021using}. Table \ref{tab1} presents the Cohen's kappa results for each code.


\begin{table}
    \setlength{\tabcolsep}{6pt} 
    \renewcommand{\arraystretch}{1.5} 
    
        \caption{Interrater reliability between automated coding process and human coding process.}\label{tab1}
        \makebox[\linewidth]{
        \begin{tabular}[t]{ *{2}{c} }
        \hline
        \textbf{Code} & \textbf{Cohen's ($\kappa$)} \\
        \hline
        Effortful Learning &  0.23 \\
        Beyond Learning Styles &  0.77 \\
        Illusions of Mastery & 0.52 \\
        Retrieval Practice, 
        Spaced out Practice, Interleaving &  0.36 \\
        \hline
        \end{tabular}
        }
\end{table}
\begin{table}
        \setlength{\tabcolsep}{6pt} 
        \renewcommand{\arraystretch}{1.5} 
        \caption{Interrater reliability between Automated + Human Keywords (A+HK) coding process and Human (H) coding process.}\label{tab3}
        \makebox[\linewidth]{
        \begin{tabular}[t]{ *{2}{c} }
        \hline
        \textbf{Code} & \textbf{Cohen's ($\kappa$)} \\
        \hline
        Effortful Learning &  0.70 \\
        Beyond Learning Styles &  0.81 \\
        Illusions of Mastery & 0.79 \\
        Retrieval Practice, Spaced out Practice, Interleaving &  0.79 \\
        \bottomrule
        \end{tabular}
        }
\end{table}

\begin{table}[!t]
\setlength{\tabcolsep}{2pt} 
\renewcommand{\arraystretch}{1.5} 
    \caption{Comparison of strength of connections between Automated + Human Keywords (A+HK) coding process and Human (H) coding process.}\label{tab4}
\makebox[\linewidth]{
\begin{tabular}[t]{ *{3}{c} }
\hline
\textbf{Connection}  &  \textbf{Strength (A+HK)} &  \textbf{Strength (H)} \\
\toprule
illusions and retrieval-interleave &  0.40 & 0.36\\
beyondLS and retrieval-interleave &  0.32 & 0.28\\
effort and retrieval-interleave &  0.30 & 0.27 \\
beyondLS and illusions & 0.27 & 0.18\\
effort and illusions &  0.27 & 0.28 \\
effort and beyondLS & 0.21 & 0.22 \\
\bottomrule
\end{tabular}
}
\end{table}


Table~\ref{tab1} shows the only code that had a Cohen’s kappa moderate level of agreement~\cite{shaffer2016tutorial} was the Beyond Learning Styles code. Illusion of Mastery had a weak level of agreement and the remaining codes had minimal level of agreement. In order to improve those numbers, we had asked the instructor, who manually coded the data, to provide us with keywords that we could include in the automatic process.

\subsection{Experiment 2}\label{subsec2}
After receiving the keywords from the instructor (Table~\ref{tab2}), we combined extracted keywords from LDA and keywords from instructor and generated a new well-formatted data table containing the same elements present in the data table from Experiment 1. Table~\ref{tab2} demonstrates that some of the keywords provided by the instructor were very similar to each other. In order to preserve the instructor's process, those keywords were not changed since they were used in the instructor's process of manually coding the data. 

\begin{table}[!tbh]
\centering
\setlength{\tabcolsep}{7pt} 
\renewcommand{\arraystretch}{1.5} 
\resizebox{0.95\textwidth}{!}{\begin{minipage}{\textwidth}
\caption{Keywords provided by the instructor.}\label{tab2}
\begin{tabular}{| m{3cm} | m{8cm}|}
\hline
\textbf{Code}  &  \textbf{Keywords} \\
\hline
Effortful Learning &  difficult, difficulties, mistakes, failure, effortful learning, desirable difficulty, desirable, effortful \\
\hline
Beyond Learning Styles &  instructional style, learning styles \\
\hline
Illusions of Mastery & illusion of mastery, illusions of mastery, misunderstanding, illusion of knowing, illusions of knowing, illusion of learning, illusions of learning, re read, cram \\
\hline
Retrieval Practice, Spaced out Practice, Interleaving &  retrieval practice, retrieval process, testing effect, test effect, recall knowledge, retrieval, actively retrieving, periodically testing, retrieval activity, retrieval activities, low stakes, effective retrieval must be repeated, flash cards, quizzing, practice and retrieval, quiz over time, continually retrieve the information, frequently quizzing, retrieval practice activity, retrieval practice activities, testing efforts, active retrieval, practice, testing for its benefit in the learning process, short quiz, active recall, process of retrieval, practice sessions, self testing, recall the information, RPA, RPAs, spacing out, spacing out practice, spaced practice, spacing practice, spaced out practice, spaced out, spaced retrieval, space retrieval, space practice, retrieval spaced, retrieve spaced, spaced application, spaced knowledge, space knowledge, spaced retrieval, retrieval practice is spaced, interleaving, interleaved practice, interleave, interleaved \\
\hline
\end{tabular}
\end{minipage}}
\end{table}

Table~\ref{tab3} presents the Cohen's kappa for the interrater reliability between the automated process with instructor's provided keywords and the human coding. Compared to simply using the automated extracted codes, the level of agreement increased. Effortful learning and Retrieval Practice, Spaced out Practice, Interleaving codes, which previously had minimal level of agreement increased to a moderate level of agreement. Illusion of Mastery which had a weak level also increased to a moderate level, and Beyond Learning Styles, which had a moderate level increased to a strong level of agreement.


\subsection{Experiment 3}\label{subsec3}
The third experiment consisted of using the well-formatted data table produced from Experiment 2 and the well-formatted data table provided by \cite{moraes2021using} to create a joint well-formatted table that included an additional column, named source, to generate the ENA visualizations using the ENA webtool. All rows that contained data generated by the algorithm were labeled "algorithm" for the source column and all rows that contained human manual coding were labeled "human" for the source column.

We used the four codes produced by our approach described in Section~\ref{sub_lda}. These codes were validated with the instructor to represent the concepts that the students were learning and therefore the concepts that the students should've connected in that online discussion. Those codes were Effortful Learning (represented simply as effort in ENA),  Beyond Learning Styles (represented as beyondLS in ENA), Illusions of Mastery (represented as illusions in ENA), and Retrieval Practice, Spaced Out Practice, Interleaving (represented as retrieval-interleaving in ENA). As described in Experiment 2 and in Table~\ref{tab3}, the interrater reliability using Cohen's Kappa reached at least $\kappa = 0.70$ for all the codes. 

As we were interested in the individual student’s network of concepts, both units of analysis and stanzas were students (i.e., all student messages) with an infinite stanza window. That configuration enabled us to visualize the connections between the codes for each student. To compare the model generated by the algorithm and the model generated by the human coder, the source column from our well-formatted data table was used.


\section{Results} \label{results}

We analyzed the results produced by each data table to detect similarities and differences between the two models generated. After that, we had a meeting with the instructor to present the results to them and evaluate the two models produced. In this section we present the results from ENA generated using data from Experiment 3 and the evaluation process conducted by the instructor.

\subsection{ENA Models} \label{ENA_results}
\begin{figure}[!tbp]
    \centering
    \subfloat[ENA using data from the A+HK coding process.]{\includegraphics[width=0.4\textwidth]{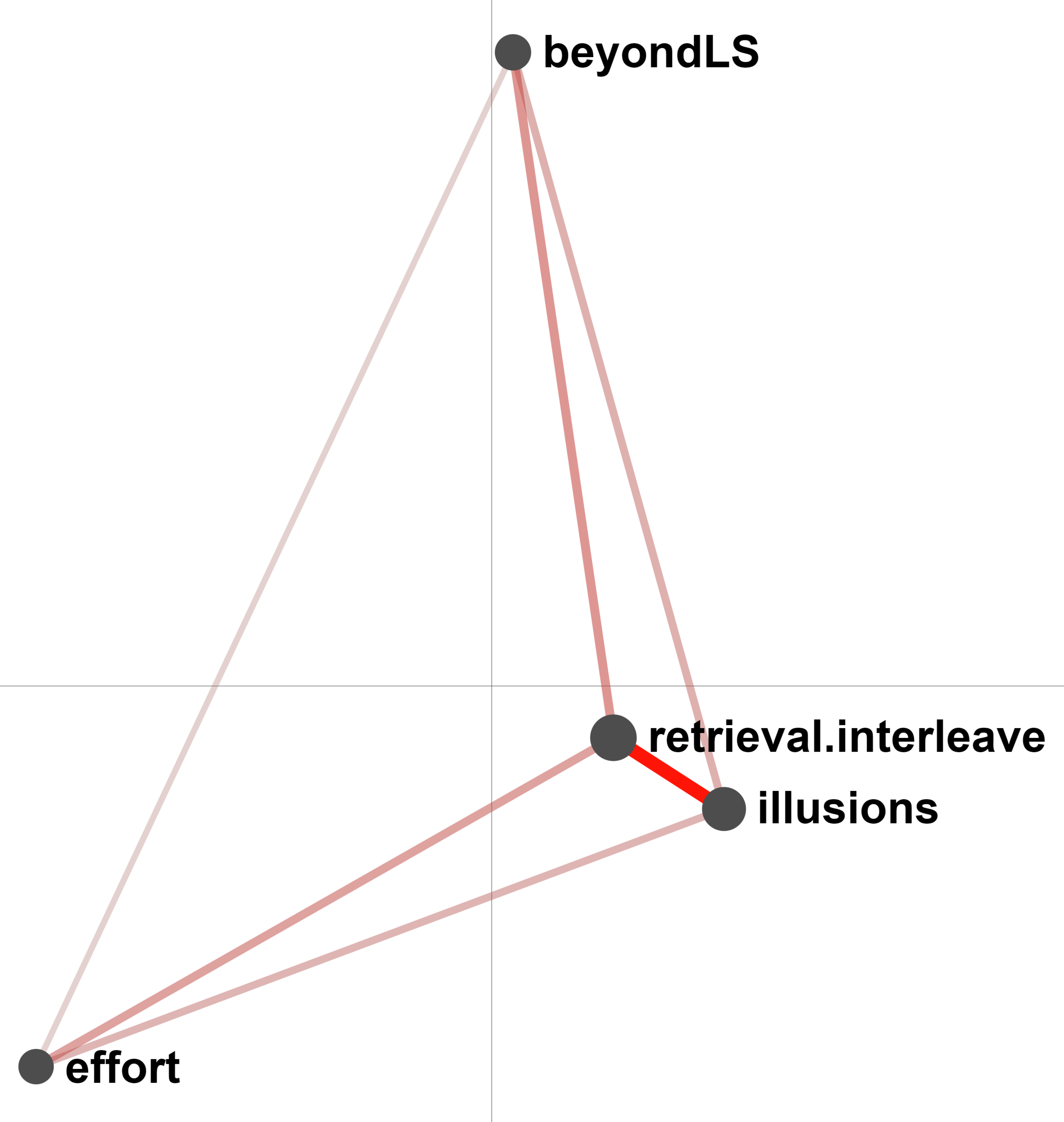}\label{fig:f1}}
    \hfill
    \subfloat[ENA using data from the H coding process.]{\includegraphics[width=0.4\textwidth]{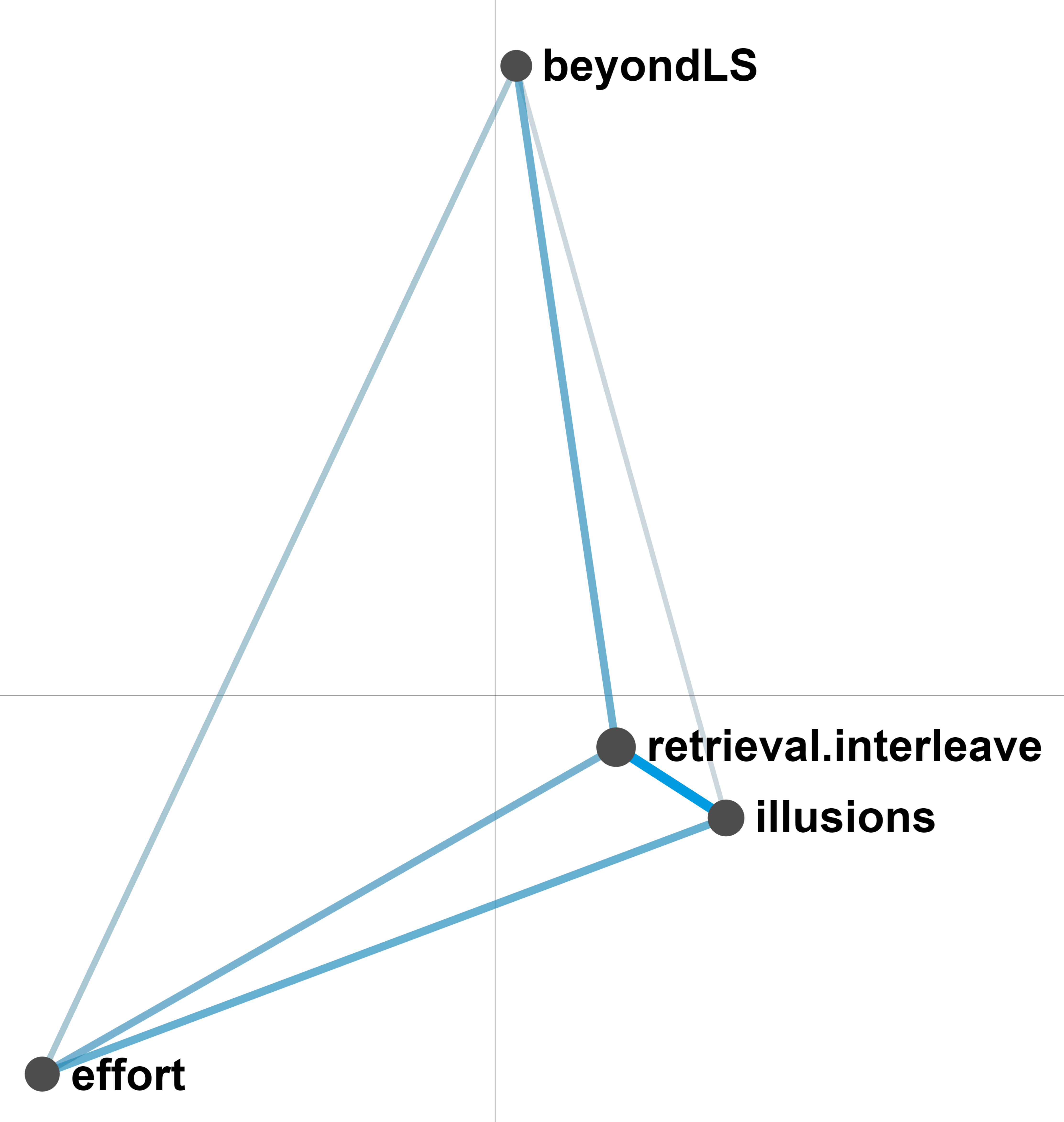}\label{fig:f2}}
    \caption{ENA Visualizations.}
\end{figure}

Figure~\ref{fig:f1} presents the group average network graph created using data from the automatic coding + human keywords (A+HK) process. The thickness of the lines between the codes indicates the strength of connections. Thicker lines indicate stronger connections, whereas thinner lines indicate weaker connections. The results indicated that for the A+HK process the strongest relationship was between the codes illusions and retrieval-interleave, followed by beyondLS and retrieval-interleave, and effort and retrieval-interleave. BeyondLS and illusions and effort and illusions had the same strength in relationship. The weakest relationship was between effort and beyondLS, as shown in Table ~\ref{tab4}.

Figure~\ref{fig:f2} presents the group average network graph created using data from the human coding process. Results show that the strongest relationships were between illusions and retrieval-interleave, followed by beyondLS and retrieval-interleave and effort and illusions. After that, the strongest relationships were between effort and retrieval-interleave and beyondLS and effort. BeyondLS and illusions connection had the weakest relation as we can observe from Table~\ref{tab4}.


\hfill
\begin{figure}
\centering
\includegraphics[width=0.4\textwidth]{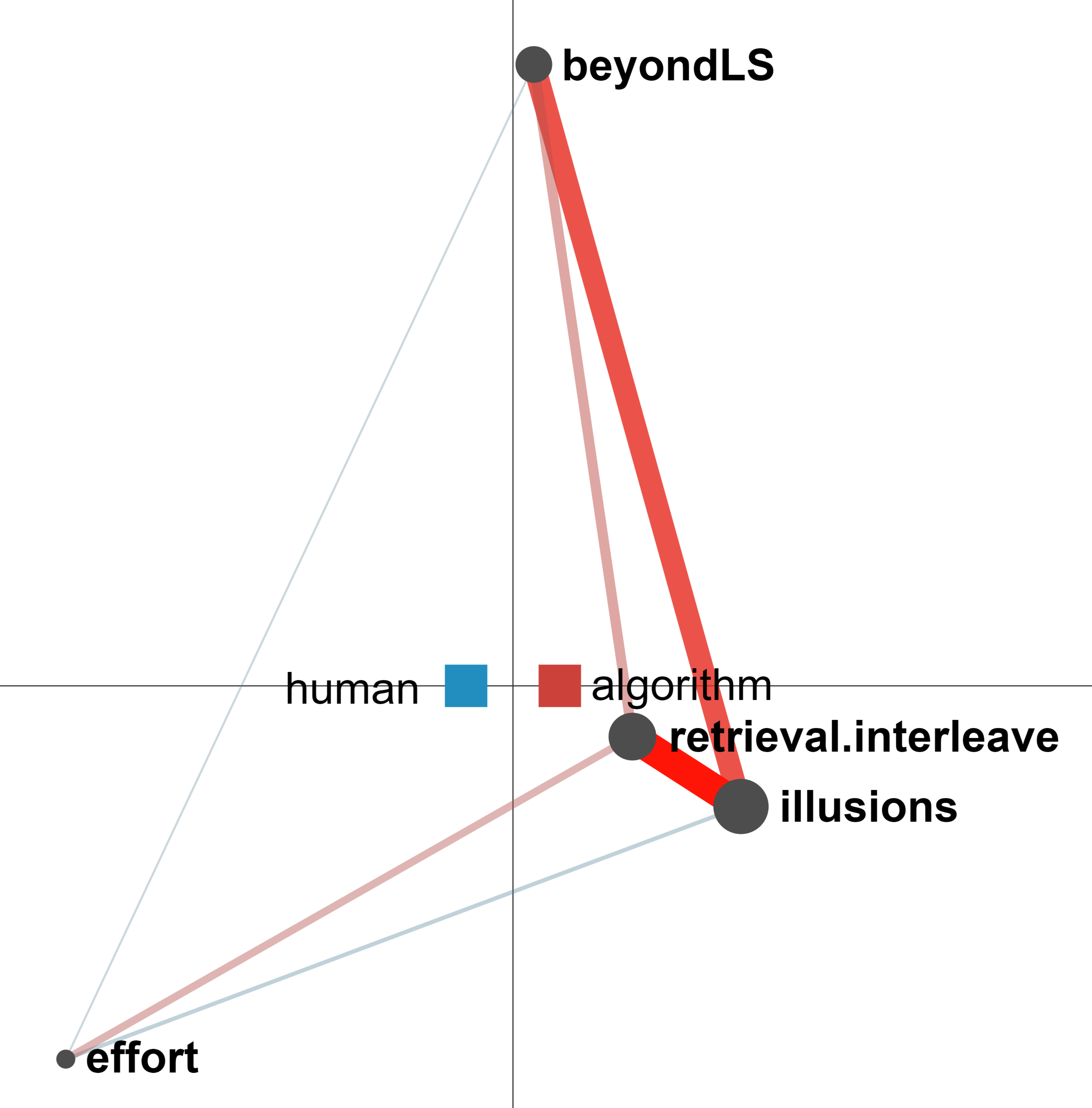}
\caption{Difference between ENA generated by the A+HK coding process and the H coding process.} \label{humanalgoENA}
\end{figure}

As shown by Figures~\ref{fig:f1},~\ref{fig:f2}, and Table~\ref{tab4}, both coding processes generated the same relationships between all the codes, with some differences between the strengths in the relationships. Figure~\ref{humanalgoENA} shows the difference between the A+HK model (named algorithm in the figure) and the human model, meaning that the A+HK made stronger connections between illusions and retrieval-interleave, beyondLS and retrieval-interleave, effort and retrieval-interleave, and beyondLS and illusions codes.

Using the ENA webtool, we performed a statistical analysis to verify that the difference between the two models was significant. Along the X axis (MR1), a Mann-Whitney test showed that Human ($Mdn=-0.13, N=25$) was not statistically significantly different at the $\alpha=0.05$ level from algorithm ($Mdn=0.13, N=25, U=206.00, p=0.04, r=0.34$). Along the Y axis (SVD2), a Mann-Whitney test showed that Human ($Mdn=-0.01, N=25$) was not statistically significantly different at the $\alpha=0.05$ level from algorithm ($Mdn=-0.01, N=25, U=318.00, p=0.92, r=-0.02$). Therefore, there is no statistical difference between the model generated by the A+HK  process and the human manual coding process. 

\begin{figure}[!tbp]
    \centering
    \subfloat[User 142854 using A+HK coding process.]{\includegraphics[width=0.38\textwidth]{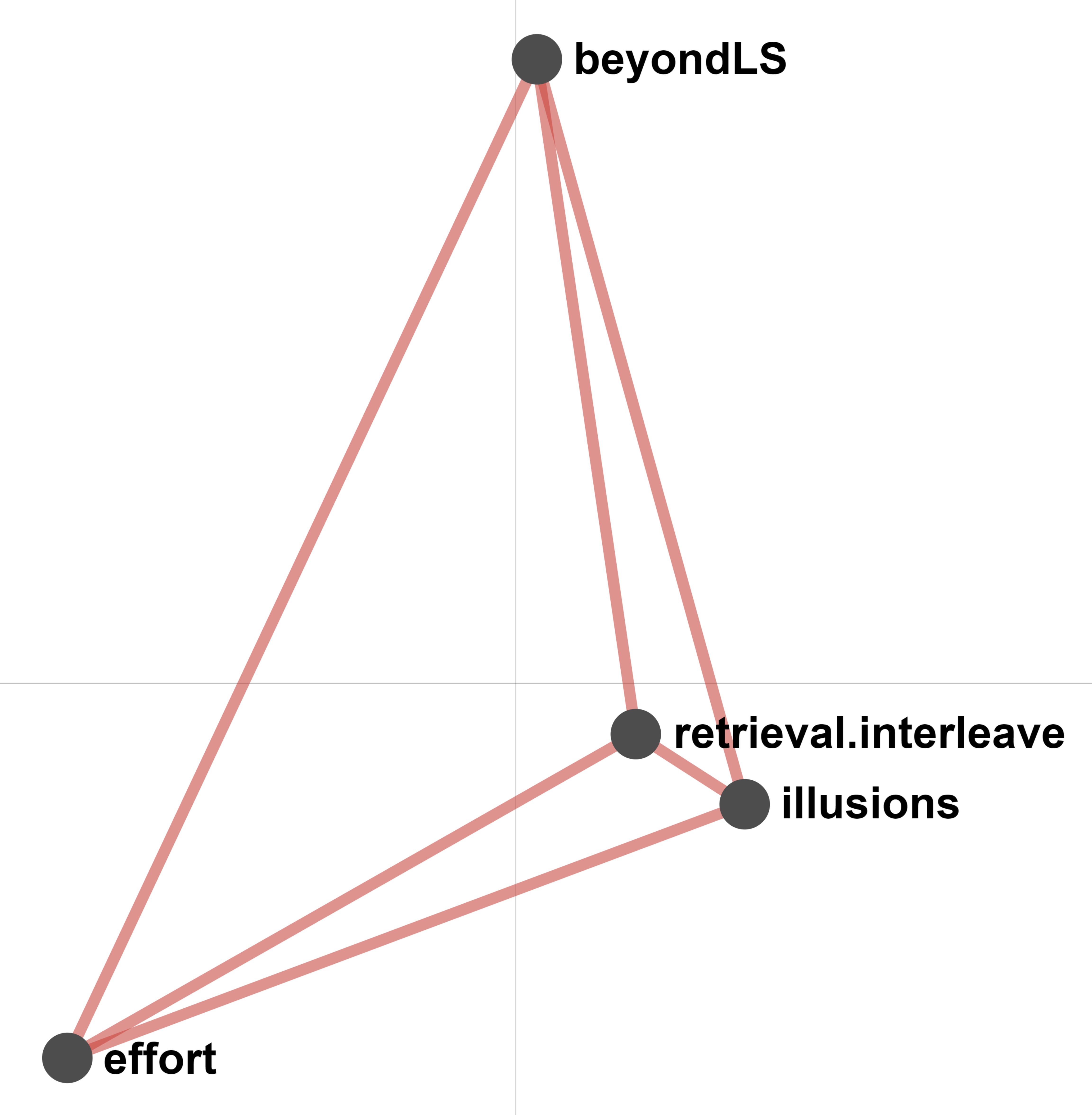}\label{fig1:f1}}
    \hfill
    \subfloat[User 142854 using H  coding process.]{\includegraphics[width=0.38\textwidth]{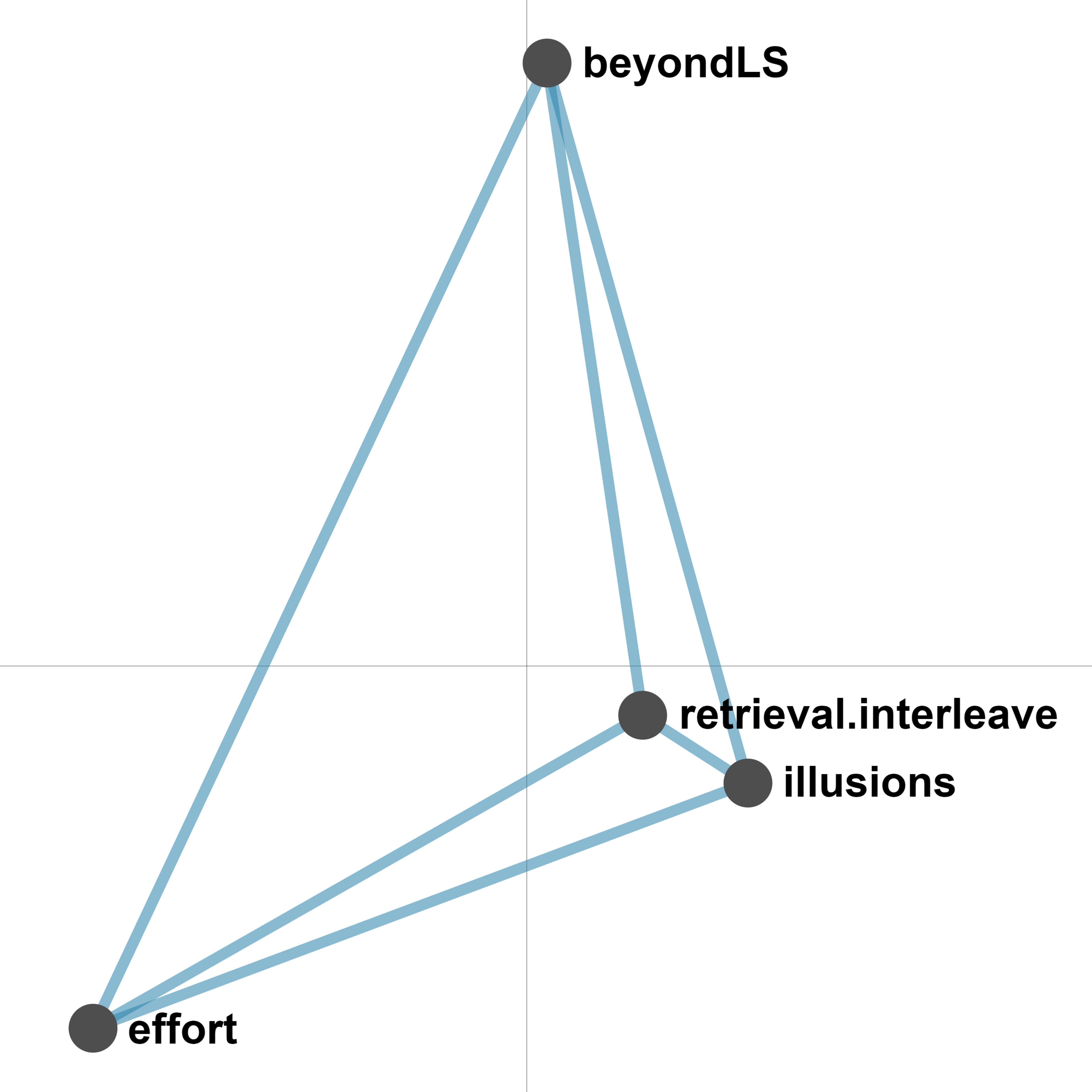}\label{fig1:f2}}
    \caption{ENA visualizations with the same structure and strengths.}
\end{figure}

\begin{figure}[!tbp]
    \centering
    \subfloat[ENA for user 135030 using A+HK coding process.]{\includegraphics[width=0.45\textwidth]{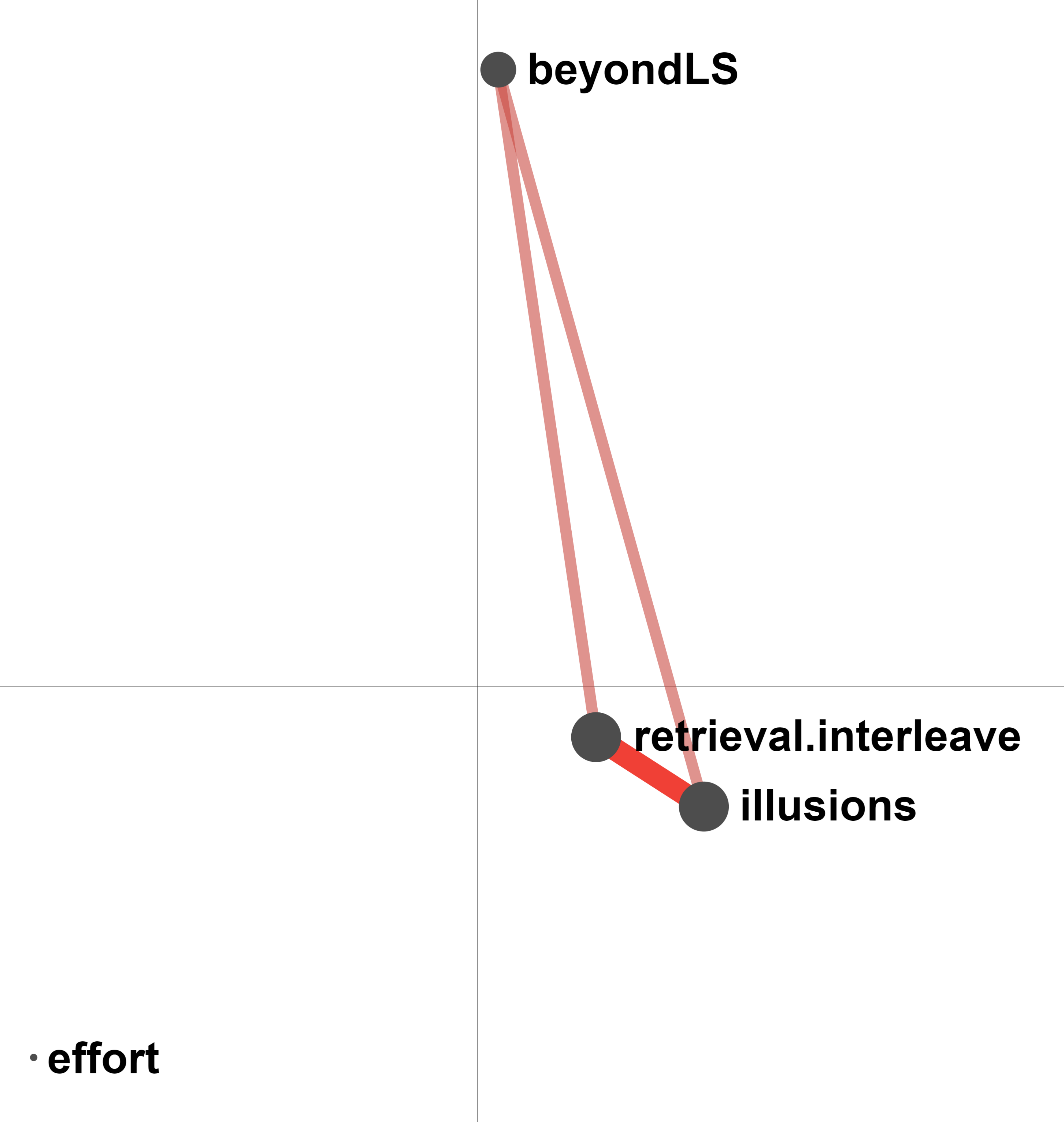}\label{fig2:f1}}
    \hfill
    \subfloat[ENA for user 135030 using H coding process.]{\includegraphics[width=0.45\textwidth]{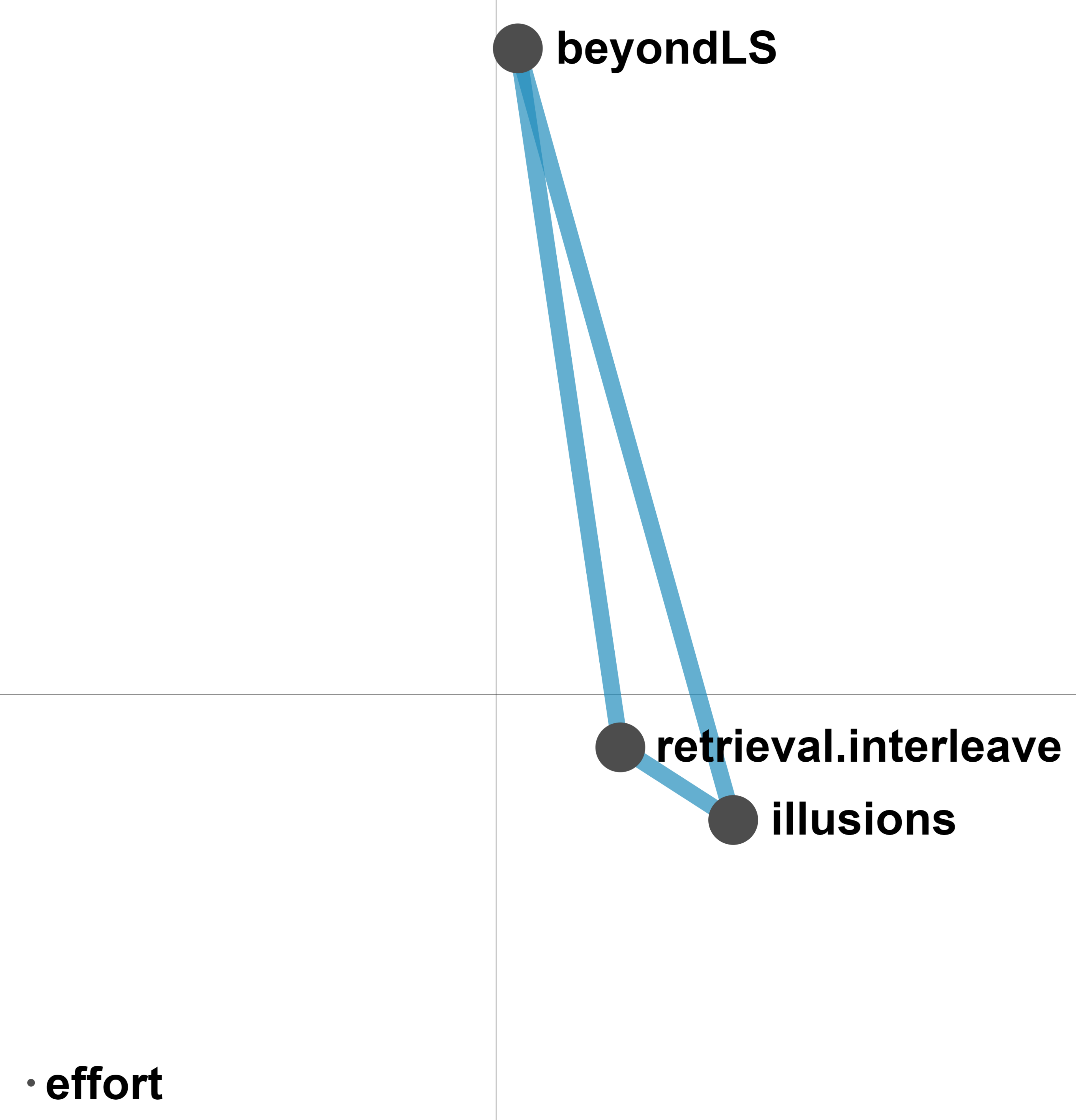}\label{fig2:f2}}
    \hfill
    \subfloat[ENA for user 125919 using A+HK coding process.]{\includegraphics[width=0.45\textwidth]{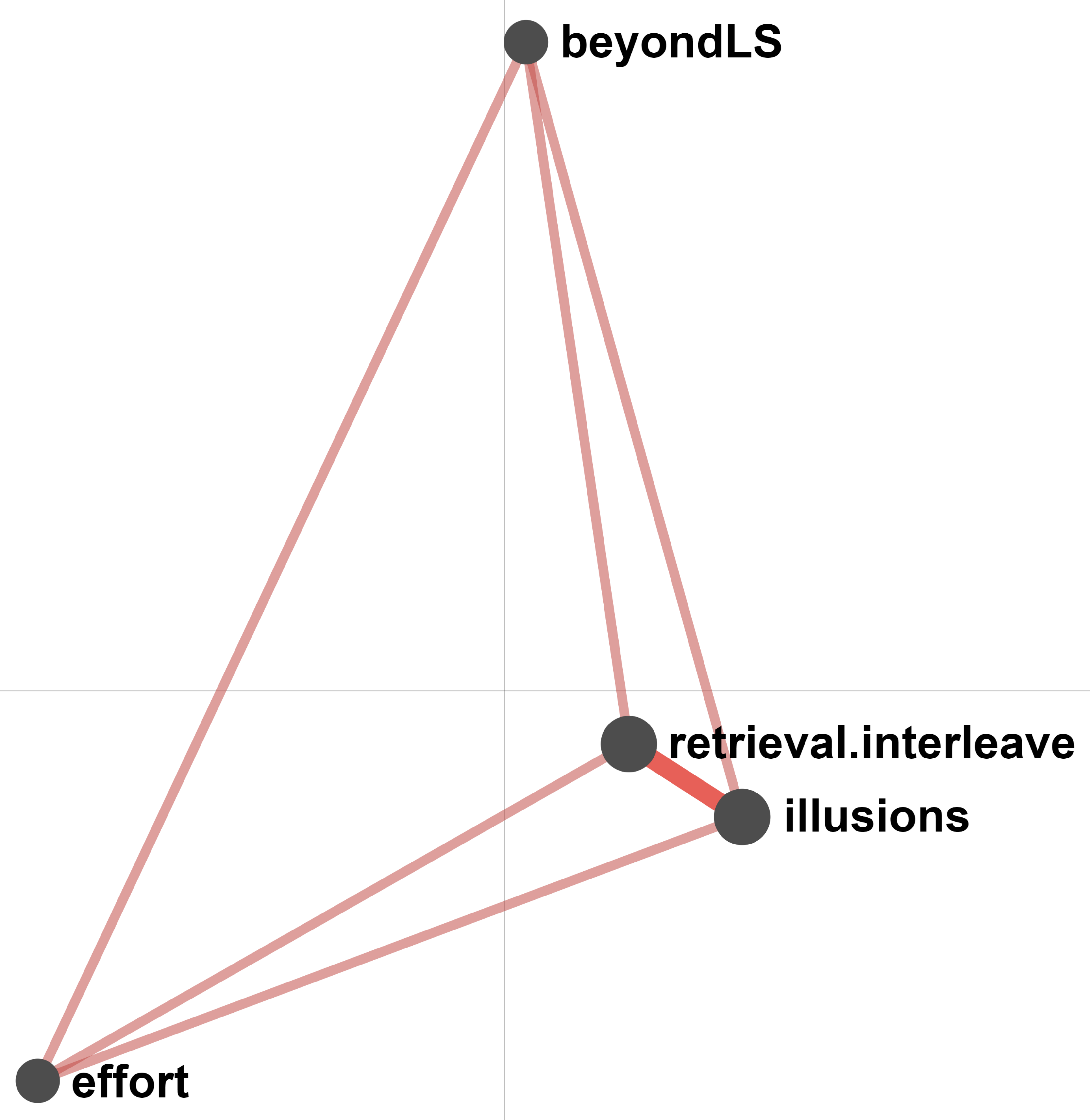}\label{fig2:f3}}
    \hfill
    \subfloat[ENA for user 125919 using H coding process.]{\includegraphics[width=0.45\textwidth]{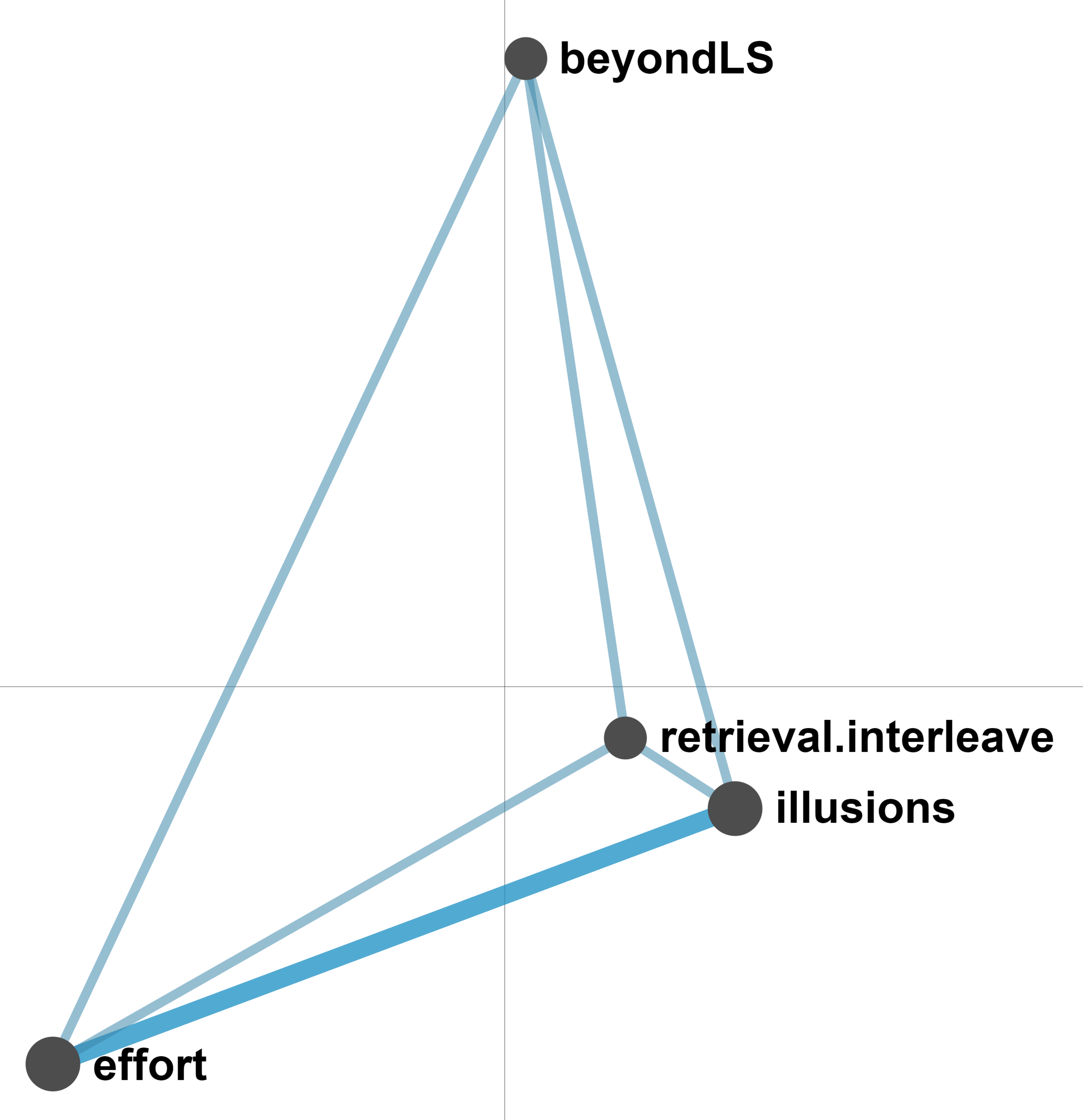}\label{fig2:f4}}
    \caption{ENA visualizations with the same structure and different strengths.}
\end{figure}

Out of the 25 ENA visualizations generated, 12 of those (48\%) had the same structure for both the A+HK process and human model. From those 12, 10 had the exact same strength in connections. One example can be seen in Figures \ref{fig1:f1} and \ref{fig1:f2}. The remaining two had different strengths. In one of those two visualizations, the structure was the same but the A+HK process found stronger connections between retrieval-interleave and illusions code and the human found stronger connections between effort and illusions instead. In the other visualization, the human coding found stronger connections than the A+HK process for all codes (Figures \ref{fig2:f1} and \ref{fig2:f2}).

The remaining 13 visualizations (52\%) had a different structure. In 10 of those, the A+HK process found more connections than the human process. In the remaining three, the human found more connections that the A+HK process.


\subsection{Instructor's Evaluation}
In order to evaluate the quality of the ENA visualizations generated using the A+HK process, we met with the course instructor. The intention was to gain feedback regarding the correctness of the models generated in cases where the automated process found connections that weren't supported by human analysis, as well as cases where the automated process found relationships that were missed by the human coder. We presented the results described on Section \ref{ENA_results} to the instructor who's familiar with Quantitative Ethnography and the ENA Web tool. Each one of the 25 visualizations were walked through with the instructor using the ENA Web tool. 

First, all ENAs that had the same structure were analyzed. All the qualitative data extracts used to find the connections between the codes for the A+HK process and the human process were looked through. As expected, for the visualizations that had the same structure and same strengths of connections, both processes used the exact same data. For those two that had a different strength between connections, in one of them (125919 user as presented in \ref{fig2:f3}) the A+HK process found a relation that the human had missed, for the other (135030 user as presented in \ref{fig2:f1}), the A+HK process presented a false positive between retrieval-interleave and illusions codes. The automated process produced a false positive in this case because it identified a keyword. However, simply having that keyword present was insufficient for the human coder to establish a relationship between the two codes.

The next step was to analyze those 13 visualizations that had different structures between the A+HK and human processes. We started by analyzing the visualizations where the A+HK process found more connections between the codes than the human process. Out of those 10 cases, in seven cases the A+HK process found connections between codes that the human had missed. Only in three cases the A+HK process generated false positives connections. To evaluate the three cases where the A+HK process found less connections, we analyzed all the qualitative data and confirmed that the A+HK process missed those connections.

\section{Discussion} \label{discussion}

\begin{itemize}\item \emph{\textbf{RQ1.} What is/are the difference(s) between the ENA visualizations generated using an automated coding process and a human coding process for the data presented in \cite{moraes2021using}?}
\end{itemize}
As we can observe from Section \ref{ENA_results} both the A+HK and the human processes generated the same structure for their network, with a small difference in the strength of some connections. The A+HK process found stronger connections between illusions and retrieval-interleave, beyondLS and retrieval-interleave, effort and retrieval-interleave, and beyondLS and illusions codes (Figure~\ref{humanalgoENA}). After running a statistical analysis, we observed that there was no statistical difference between the model generated by the A+HK and the human. This could potentially be a good indicator that an automated process that used combined LDA keywords and human keywords can contribute to generating ENA visualizations to help instructors in evaluating asynchronous online discussion data. Further tests need to be done in other sections of the course to confirm that similar structures will be found between the A+HK and human process for those new datasets.

\begin{itemize}\item \emph{\textbf{RQ2.} How does an instructor evaluate the ENA visualizations generated?}
\end{itemize}
During the evaluation process, the instructor pointed out that the course used a series of assignments that required students to synthesize concepts into coherent discussion posts. Consequently, grading those posts demanded frequent reading to identify concepts and to evaluate how well each student integrated them. Recognizing the grading challenges, the instructor considered the potential usefulness of an algorithm generated ENA for potentially improving the efficiency and accuracy of grading.
Comparing the A+HK coding to the human coded equivalent, the instructor mentioned that it was impressive that the A+HK generated identical coding for 11 of the 25 students. Additionally, they noted it was encouraging to see that the algorithm found more connections in 11 additional cases. In other words, 22 out of the 25 cases (88\%) the A+HK identified the correct conceptual connections in the written passages and identified more correct connections in 11 out of the 25 written passages (44\%). The level of accuracy was promising, suggesting that it may be possible to use the A+HK to highlight the majority of the connections automatically.

However, the A+HK method found fewer relationships in three cases (12\%). Examining those cases, the human coder identified accurate relationships, but those relationships were extrapolated from the subtle meaning and content in the post. Future work needs to be done on how to include those aspects in the automated process. For example, improving the keywords offered by the instructor and using Large Language Models to consider the context of words ~\cite{ghaffari2023grounding} for a stronger connection between codes discussed in a course.

\section{Conclusion} \label{conclusion}
In this paper we presented an approach that clustered topic keywords into meaningful categories from a relatively small online course discussion dataset using Latent Dirichlet Allocation (LDA) \cite{blei2003latent}. Those keywords and the instructor's keywords were then used to automatically code asynchronous online discussion data. Finally, ENA visualizations were generated based on the data. The visualizations were compared with the corresponding visualizations generated by human coding process, and both visualizations were evaluated by an instructor. Results indicated that there is no statistical difference between the model generated by the A+HK process and the human. 

Overall, the result of the A+HK demonstrates significant potential to assist instructors in evaluating discussion based assignments that demand the students' synthesis and integration of concepts, especially in larger classrooms. An automated method allows instructors teaching classes with hundreds of students to use discussion posts to promote these higher order learning outcomes. It is important to acknowledge that our approach is considered as a tool for instructors to enhance their evaluation process of asynchronous online discussions. Additional efforts need to be made in order to verify its applicability to other class settings such as different student populations and different course materials.

\bibliographystyle{splncs04}
\bibliography{references}
\end{document}